# Searching the hearts of graphene-like molecules for simplicity, sensitivity and logic

Sara Sangtarash,‡[*a] Cancan Huang,‡[b] Hatef Sadeghi,‡[a] Gleb Sorohhov,[b] Jürg Hauser,[b] Thomas Wandlowski,[b] Wenjing Hong,[*b] Silvio Decurtins,[b] Shi-Xia Liu[*b] and Colin J. Lambert[*a]

[a] Quantum Technology Centre, Lancaster University, Lancaster LA1 4YB, UK.
[b] Department of Chemistry and Biochemistry, University of Bern, Freiestrasse 3, CH-3012 Bern, Switzerland.
* s.sangtarash@lancaster.ac.uk; c.lambert@lancaster.ac.uk; hong@dcb.unibe.ch; liu@dcb.unibe.ch.
‡ These authors contributed equally to this work.

**ABSTRACT:** If quantum interference patterns in the hearts of polycyclic aromatic hydrocarbons (PAHs) could be isolated and manipulated, then a significant step towards realizing the potential of single-molecule electronics would be achieved. Here we demonstrate experimentally and theoretically that a simple, parameter-free, analytic theory of interference patterns evaluated at the mid-point of the HOMO-LUMO gap (referred to as M-functions) correctly predicts conductance ratios of molecules with pyrene, naphthalene, anthracene, anthanthrene or azulene hearts. M-functions provide new design strategies for identifying molecules with phase-coherent logic functions and enhancing the sensitivity of molecular-scale interferometers.

## INTRODUCTION

Single-molecule electronic junctions are of interest not only for their potential to deliver logic gates, sensors and memories with ultra-low power requirements and sub-10 nm device footprints, but also for their ability to probe room-temperature quantum properties at a molecular scale. For example, when a single molecule is attached to metallic electrodes, de Broglie waves of electrons entering the molecule from one electrode and leaving through the other form complex interference patterns inside the molecule.[1-3] Nowadays there is intense interest in utilising these patterns in the optimisation of single-molecule device performance. Indeed, electrons passing through single molecules have been demonstrated to remain phase coherent, even at room temperature[3-5] and a series of theoretical and experimental studies have shown that their room-temperature electrical conductance is influenced by quantum interference (QI).[6-19]

In practice, the task of identifying and harnessing quantum effects is hampered, because transport properties are strongly affected by the method used to anchor single molecules to electrodes[20-30]. This makes it difficult to identify simple design rules for optimising the electronic properties of single molecules. Furthermore few analytic formulae are available, which means that pre-screening of molecules often requires expensive numerical simulations. In what follows, our aim is to introduce a new concept for elucidating QI patterns within the hearts of molecules, caused by electrons entering the molecule with energies $E$ near the mid-point of the HOMO-LUMO (H-L) gap. We refer to these mid-gap interference patterns as M-functions. The approach is intuitive and leads to a simple, parameter-free, analytical description of molecules with polycyclic aromatic hydrocarbon (PAH) cores, which agrees with experiment to an accuracy comparable with ab initio calculations.

A typical single-molecule junction involves metallic electrodes, connected via linker groups to the heart (*ie* core) of the molecule. Fig. 1 shows two such molecules, with a common pyrene-based heart, connected by acetylene linkers to gold electrodes. Such PAHs are attractive for molecular electronics,[31-35] because they are defect free and provide model systems for electron transport in graphene, treated as an infinite alternant PAH.[36,37] As part of our demonstration of the utility of M-functions, we present mechanically-controlled break-junction (MCBJ) measurements of the electrical conductance of these molecules. **P1** and **P2** are examples of molecules with identical hearts, but different connectivities. **P1** is connected to acetylene linker groups at positions labelled 2 and 9, whereas **P2** is connected at positions 3 and 10.

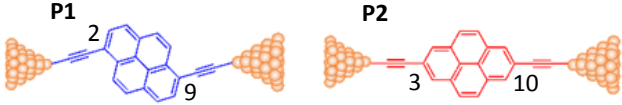

**Fig. 1** Two molecules **P1** and **P2** with common pyrene hearts, but different connectivities to gold electrodes. (See figure 2 for a more detailed numbering convention used in this study. This does not correspond to the usual chemical numbering convention, but it is convenient analytically and allows us to assign labels to all atoms.)

The amplitude of the interference pattern on an atomic orbital *i* due to an electron of energy *E* entering a core at orbital *j* will be denoted by the M-function $M_{i,j}(E)$. In what follows, it will be convenient to introduce the dimensionless energy $E_M$, which measures the electron energy *E* relative to the middle of the H-L gap $E_{HL}$, in units of the half width of the H-L gap. If $E_H$ ($E_L$) is the energy of the HOMO (LUMO) of the core of the molecule, we define the dimensionless energy $E_M = (E - E_{HL})/\delta_{HL}$, where $\delta_{HL} = (E_H - E_L)/2$ and $E_{HL} = (E_H + E_L)/2$. For PAHs represented by bipartite lattices possessing a symmetric energy spectrum and a filled HOMO, the mid-gap energy $E_{HL} = 0$ lies at the centre of the spectrum and the mid-gap interference patterns obey simple rules[3]. More generally, mid-gap transport involves interference at finite $E_{HL}$, and therefore in what follows, we generalise these rules to encompass interference patterns at all energies within the gap. As shown below, this distinction is particularly important for non-symmetric molecules such as azulene, for which conventional rules for quantum interference break down[38]. When *E* is close to $E_H$ or $E_L$ (ie $E_M = \pm 1$) a Breit-Wigner description based on a HOMO or a LUMO resonance is relevant, and therefore one might be tempted to suppose that near the mid-gap, a description based on a superposition of HOMO and LUMO levels would suffice. Such a description would not be accurate, because at $E_M = 0$, states such as HOMO-1, LUMO+1, etc make comparable contributions. From the viewpoint of mid-gap quantum transport, such resonances are a distraction and therefore M-functions are defined such that these irrelevancies are removed.

**ANALYTIC FORMULAE FOR M-FUNCTIONS**

Mathematically we define $M_{i,j}(E) = D(E)G_{i,j}(E)$, where $G_{i,j}(E)$ is the *i,j*th element of the Green's function $G(E) = (E - H)^{-1}$ of the Hamiltonian *H* describing the isolated core and $D(E)$ is a function chosen to cancel divergencies of $G(E)$, which arise when *E* coincides with an eigenvalue of *H*. In the absence of degeneracies, it is convenient to choose $D(E)$ to be proportional to the determinant of *(E-H)*. (See SI for more details, along with a list of M-function properties.)

In what follows we shall construct an intuitive description of mid-gap transport, which in its simplest form is parameter-free and describes how connectivity alone can be used to predict the interference patterns created by electrons of energy $E_{HL}$ passing through the heart of PAHs. When linker groups, which are weakly coupled to orbitals *i* and *j*, are in contact with metallic electrodes whose Fermi energy $E_F$ lies at the mid-gap $E_{HL}$, the resulting (low-temperature) electrical conductance $\sigma_{i,j}$ is proportional to $[M_{i,j}(E_{HL})]^2$.[1,3] Therefore the ratio of two such conductances (associated with links *i* and *j* or *l* and *m*) is given by the mid-gap ratio rule (MRR):

$$\sigma_{i,j}/\sigma_{l,m} = \left[M_{i,j}(E_{HL})/M_{l,m}(E_{HL})\right]^2 \qquad (1)$$

In what follows, we report MCBJ measurements of the conductances of molecules **P1** and **P2** and show that their statistically-most-probable conductances obey the MRR. We also demonstrate that the MRR agrees with literature measurements of molecules with naphthalene, anthracene, anthanthrene and azulene hearts. This is a remarkable result, since M-functions and the mid-gap energy $E_{HL}$ contain no information about the electrodes. This agreement between experiment and the MRR is evidence that in these experiments, the Fermi energy of the electrodes lies close to the mid-gap energy. Having demonstrated the predictive nature of M-functions, we further discuss their utility by showing that M-functions lead to new design strategies for identifying phase-coherent logic functions, and for increasing the sensitivity of molecular-scale interferometers.

In general, M-functions depend on the parameters describing the underlying Hamiltonian *H* of the core. However, for the purpose of calculating the contribution to interference patterns from π-orbitals, graphene-like cores can be represented by lattices of identical sites with identical couplings, whose Hamiltonian *H* is simply proportional to a parameter-free connectivity matrix *C*. In this case, for electrons of energy *E* entering the core at site *i* and exiting at site *j*, the M-function $M_{ij}(E)$ is also parameter-free and depends on connectivity alone.

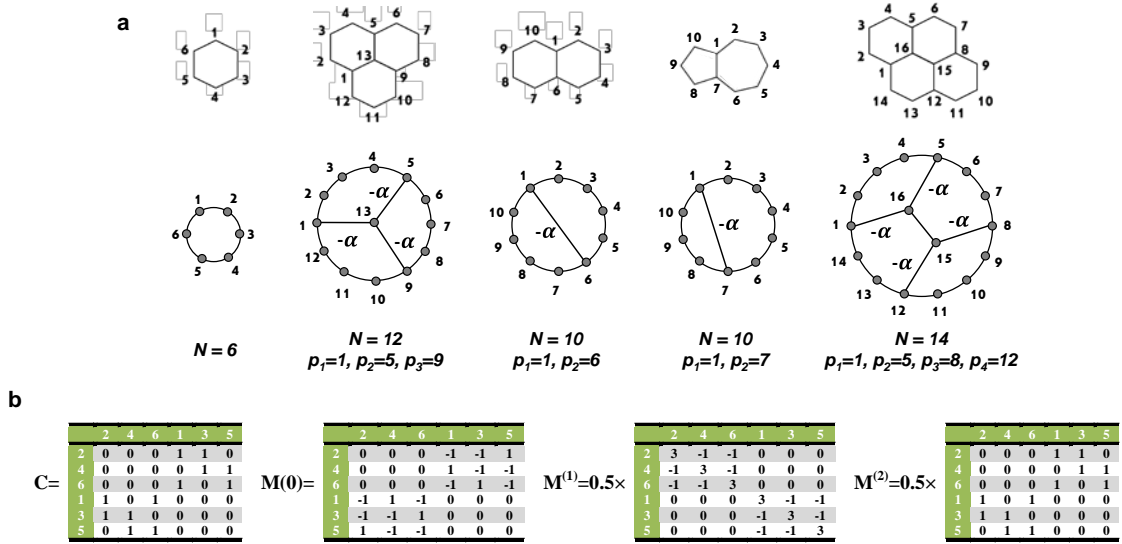

**Fig. 2** Examples of bipartite and non-bipartite molecules and their lattice representation. The upper row of (a) shows lattice representations of a benzene ring, a phenalenylium cation, naphthalene, azulene and pyrene. The lower row shows connectivity-equivalent representations of these lattices, which can be viewed as rings of peripheral sites, perturbed by the presence of additional bonds to sites p1, p2, etc. Notice that all lattices except azulene are bipartite (ie odd-numbered sites are connected to even-numbered sites only). (b) The connectivity table $C$ of a benzene ring and all contributions to eq. (3) for the benzene M-function. Notice that $C$ is block off-diagonal and as a consequence, (see eqs (42) and (46) of the SI), $M(0)$, $M^{(2)}$ are also block off-diagonal, whereas $M^{(1)}$ is block diagonal.

As an example, consider a ring of $N$ sites, labelled by integers, which increase sequentially in a clockwise direction, as shown in figure 2. For a benzene ring (where $N=6$), figure 2b shows the corresponding connectivity table $C$, obtained by placing a '1' at all entries for which a connection exists between neighbouring sites in the ring. In the simplest π-orbital description of such a ring, where neighbouring sites are connected by couplings ($-\gamma$), the Hamiltonian $H$ is related to the connectivity matrix $C$ by $H = -\gamma C$ and as discussed in the SI, the M-function of the ring is given by[1,3]

$$M_{ij}(E) = \cos k(|j-i| - N/2) \quad (2)$$

where $k(E) = \cos^{-1}[-E/2\gamma]$. Without loss of generality, the parameter $\gamma$ will be set to unity, because it cancels in the MRR, yielding a parameter-free theory. In this case, $E_H = -1$, $E_L = 1$, $E_{HL}=0$ and the dimensionless energy is $E_M = E$. For a given value of $E$, the numbers $M_{ij}(E)$ form a table of an energy-dependent functions, which we call an M-table $M(E)$. For $N=6$, there are four distinct entries, namely $M_{ii}(E) = \cos 3k = \frac{3E_M}{2}(1 - \frac{E_M^2}{3})$, $M_{i,i+1}(E) = \cos 2k = \frac{E_M^2}{2} - 1$, $M_{i,i+2}(E) = \cos k = -\frac{E_M}{2}$ and $M_{i,i+3}(E) = 1$. At $E = E_{HL} = 0$, as expected, this table reveals that the π-orbital contribution to the electrical conductance of meta-connected cores such as $i=1$ and $j=3$ is zero, whereas the conductances of para ($i=1$ and $j=4$) and ortho ($i=1$ and $j=2$) connected cores have the same non-zero conductance. In other words the conductance ratio $[M_{13}(E_{HL})/M_{14}(E_{HL})]^2$ vanishes, whereas the ratio $[M_{12}(E_{HL})/M_{14}(E_{HL})]^2 = 1$. On the other hand, if $E$ is allowed to vary relative to the H-L gap centre, then these ratios change. This example illustrates M-function property (see SI for a list of M-function properties) that M-functions can be represented by low-order polynomials in $E_M$, in contrast with Green's functions, which are non-analytic and require infinite power series. Indeed, the above expressions can be written

$$M(E) = M(E_{HL}) + M^{(1)} E_M + M^{(2)} E_M^2 + M^{(3)} E_M^3 \quad (3)$$

where $E_{HL} = 0$, $M^{(3)} = -1/2\, I$ (with I the unit matrix) and $M(E_{HL})$, $M^{(1)}$ and $M^{(2)}$ are shown in figure 2b. This result illustrates another general property of M-functions (see SI), namely that the low-order M-tables $M^{(1)}$, $M^{(2)}$, etc can be constructed from a knowledge of $M(E_{HL})$ alone. For example for benzene, the general relationship (see SI) between these tables reduces to $M^{(1)} = \frac{1}{2} M^2(E_{HL})$ and $M^{(2)} = M(E_{HL})\{\frac{1}{4}[M^2(E_{HL}) - 5]\}$, as can be checked by direct multiplication of $M(E_{HL})$ in figure 2b. This means that interference patterns at energies $E$ in the vicinity of the mid-gap can be generated solely from the mid-gap interference patterns $M_{ij}(E_{HL})$.

Equation (2) demonstrates that quantum interference rules established for mid-gap transport are modified when $E \neq E_{HL}$ (ie $E_M \neq 0$). For example at $E_M = 0$, where $M_{ij}(E) = M_{ij}(E_{HL})$, $M_{ij}$ vanishes if $ij$ are both even or both odd. On the other hand, for finite $E_M \neq 0$ this simple rule is invalid and instead finite-energy M-tables such as equation (3) should be used.

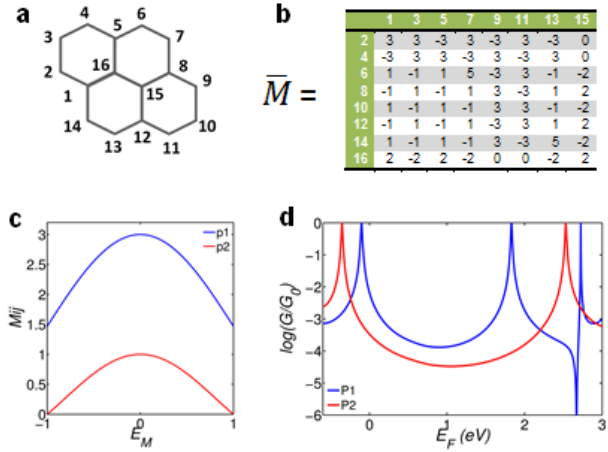

**Fig. 3** Calculated conductances of **P1** and **P2**. (a) The numbering convention for the cores of molecules **P1** and **P2** shown in fig. 1. (b) As noted in eq. (35) of the SI, the mid-gap M-table M(0) is block off-diagonal and of the form $M(0) = \begin{pmatrix} 0 & \bar{M}^t \\ \bar{M} & 0 \end{pmatrix}$. Fig 3b shows the off-diagonal block $\bar{M}$ of the mid-gap M-table. (c) M-functions of **P1** ($M_{2,9}$) and of **P2** ($M_{3,10}$) for energies $E$ varying between the pyrene HOMO ($E_H$) and the pyrene LUMO ($E_L$). Both functions are plotted against the dimensionless energy $E_M = (E - E_{HL})/\delta_{HL}$, where $\delta_{HL} = (E_H - E_L)/2$ and $E_{HL} = (E_H + E_L)/2$. The pyrene HOMO (LUMO) corresponds to $E_M = -1$ ($E_M = +1$). (d) NEGF results for the electrical conductance of **P1** and **P2** as a function of the Fermi energy $E_F$ of the electrodes at zero temperature.

Equation (2) is the simplest example of an M-function. It is also a useful starting point for obtaining analytic expressions for M-functions of other PAH cores, such as those shown in figure 2a, because if the bonds denoted $-\alpha$ in figure 2a are set to zero, the peripheral sites of these cores are equivalent to a ring of $N$ sites, whose M-functions are given by equation (2). On the other hand, when $\alpha = \gamma$, electrons traversing the periphery of the ring are scattered at peripheral sites $p$, labelled $p_1$, $p_2$ etc., which are connected by the bonds $\alpha$. For example for pyrene, $p_1 = 1$, $p_2 = 5$, $p_3 = 8$ and $p_4 = 12$. Analytic formulae for the resulting M-functions of these lattices are presented in the SI. For pyrene, (see equ. (25) of the SI), $E_H = -0.45\gamma$, $E_L = 0.45\gamma$, for naphthalene, $E_H = -0.62\gamma$, $E_L = 0.62\gamma$ and as expected, for such bipartite lattices the gap centre is at $E_{HL} = 0$. On the other hand, for the non-bipartite azulene (see equ. (33) of the SI), $E_H = -0.48\gamma$, $E_L = 0.4\gamma$ and the gap centre is at $E_{HL} = -0.04\gamma$. In this case, $E_M = (E + 0.04\gamma)/0.88\gamma$ and $E_M = 0$ does not coincide with $E = 0$.

Having introduced the concept of M-functions and energy-dependent M-tables, we now use these to examine mid-gap conductance ratios of molecules with either bipartite or non-bipartite PAH cores. First we examine the conductance ratios of the pyrene-based molecules **P1** and **P2** of figure 1, which possess bipartite cores and for which $E_{HL}=0$. Secondly, we compare the predictions of the MRR with literature measurements for the conductances of other molecules with both bipartite and non-bipartite hearts[38] and with DFT and GW predictions. An analytic formula for the M-functions of a pyrene heart, is derived in the SI. The resulting zero-energy M-table $M(0)$ is block off-diagonal of the form $M(E_{HL}) = M(0) = \begin{pmatrix} 0 & \bar{M}^t \\ \bar{M} & 0 \end{pmatrix}$, where $\bar{M}$ is a table of integers, as shown in figure 3b. As examples, $M_{2,9}(E)$ and $M_{3,10}(E)$ are plotted in figure 3c. These yield for **P1**, $M_{2,9}(0)= -3$ and for **P2**, $M_{3,10}(0) = -1$. Hence the MRR predicts a mid-gap conductance ratio of $\sigma_{2,9}/\sigma_{3,10} = (3/1)^2 = 9$.

**EXPERIMENTAL AND DFT RESULTS FOR PYRENE**

We now verify the above MRR prediction by measuring the electrical conductances of pyrene cores with TMS-protected (TMS = trimethylsilyl) acetylene groups at different positions, **P1**[39] and **P2**[40] using the MCBJ technique.[41] The repeating opening and breaking cycles are carried out in a solution containing 0.1 mM target molecules in a mixture of THF:TMB (mesitylene) = 1:4 (v:v). Then 0.2 mM tetrabutylammonium (TBAF) in a mixture of THF:TMB = 1:4 (v:v) solution was added for *in-situ* cleavage reaction of the TMS protection group.[42,43] Figure 1 shows the schematics of the **P1** and **P2** molecular junctions *via* the anchoring through a C-Au bond between both gold electrodes.

Figure 4a shows some typical individual stretching traces from the MCBJ measurement of **P1** and **P2** molecules. For both molecules, current-voltage traces were found to be linear. A sharp conductance decrease occurs after the rupture of gold-gold atomic contacts (plateau at conductance quantum $G_0$), followed by clear but tilted molecular plateaus for the individual traces. Based on 1000 individual traces, the conductance histograms were constructed without data selection, as shown in figure 4b. The most probable conductance of **P1** locates at $10^{-3.3\pm0.1}$ $G_0$ while the most probable conductance of **P2** is almost one order of magnitude lower at $10^{-4.2\pm0.1}$ $G_0$. Two dimensional (2D) histograms in figure 4c,d reveal that the molecular plateaus are observed in almost all stretching traces, suggesting a ~100% junction formation probability by the *in-situ* cleaving off reaction of the TMS groups, which agrees well with the previous study using the TMS cleaving-off reaction for the formation of a single-molecule junction.[42] The stretching distance distributions of the two molecules (insets of figure 4c,d) suggest a 0.2 nm difference between the two molecules, which is in good agreement with X-ray structural data, giving Si-Si separations of 14.5 Å (P1) (see SI) and 16.0 Å (P2),[40] respectively. The experimental conductance ratio of $10^{-3.3}/10^{-4.2}$ is approximately 8, which compares favourably with the MRR prediction of 9. The occurrence of tilted plateaus for both molecules suggests that during the stretching process, due to the enhanced strength of the Au–C interaction, a single gold atom is detached from the electrode surface while the gold–carbon bond does not break[44].

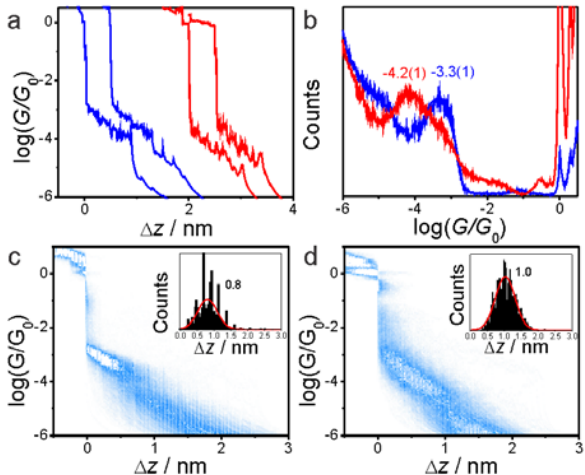

**Fig. 4** Measured conductances of **P1** and **P2**. (a) Typical conductance-relative distance traces and (b) one-dimensional (1D) conductance histograms of **P1** (blue) and **P2** (red) molecules. (c,d) Two-dimensional (2D) conductance histograms and stretching distance distributions (inset) of **P1** (c) and **P2** (d).

To further verify the MRR prediction, figure 3d shows the electrical conductances of **P1** and **P2** as a function of the Fermi energy $E_F$ of the electrodes, obtained from a transport calculation using a combination of density functional theory (DFT) and non-equilibrium Green's functions (See methods). From these results, the predicted conductance ratio varies between 10 and 7 over the range $0 < E_F < 1.2$ and achieves a value of 9 at $E_F = 0.05$, which is close to the DFT-predicted Fermi energy of $E_F = 0$.

## COMPARISON WITH LITERATURE RESULTS

Pyrene possesses a bipartite heart, in which atoms labelled by even integers are connected only to atoms labelled by odd integers and the numbers of odd and even-numbered atoms are equal. We now show that M-functions describe non-bipartite lattices such as azulene, whose M-functions have no particular symmetry and whose values at the gap centre are not integers. This molecule is a challenge, because well-known bond counting rules for predicting QI[45,46] have been shown to break down in azulene cores[38]. Four examples ($M_{8,10}(E)$, $M_{4,9}(E)$, $M_{3,6}(E)$ and $M_{3,5}(E)$) of the analytic formula (see SI) for azulene M-functions are plotted in figure S1c of the SI. These examples allow us to test the MRR against measurements of the electrical conductance of molecules with azulene cores[38], where it was reported that $\sigma_{8,10} = 32\times10^{-5}G_0$, $\sigma_{4,9} = 32\times10^{-5}G_0$, $\sigma_{3,6} = 8\times10^{-5}G_0$ and $\sigma_{3,5} = 2\times10^{-5}G_0$ yielding experimental conductance ratios of $\sigma_{4,9}/\sigma_{8,10} = 1$, $\sigma_{5,8}/\sigma_{8,10} = 1/4$ and $\sigma_{3,5}/\sigma_{8,10} = 1/16$.

Table 1 shows a comparison between MRR and experiment, and demonstrates good agreement between the experiment and our parameter free mid-gap MRR. For example, the ratio between connectivity 4,9 and 8,10 of azulene is measured to be 1, whereas the GW calculation[38] and our DFT-NEGF calculation yields a ratio of 0.32 and 0.93 respectively. These predictions were obtained by treating the Fermi energy as a free parameter and adjusting it to yield the closest agreement with experiment. For example, in the GW calculations, the Fermi energy is chosen to be far from the GW predicted Fermi energy (-1.5eV). In contrast, our parameter free MRR, which has no such freedom, predicts a ratio of 0.72, in much better agreement with the experiment. For completeness, Table 1 also shows excellent agreement between the parameter free mid-gap MRR and existing experimental values for naphthalene[15], anthracene[15], and anthanthrene[2].

The above result is remarkable, because if the Fermi energy $E_F$ of external electrodes does not coincide with the mid-gap $E_{HL}$ then the MRR should be replaced by

$$\sigma_{i,j}/\sigma_{l,m} = [M_{i,j}(E_F)/M_{l,m}(E_F)]^2. \qquad (4)$$

The fact that the MRR agrees with experiment suggests that in all of the above measurements, $E_F$ is close to the mid-gap.

**Table 1** The top three rows show a comparison between the mid-gap MRR, GW and experimental conductance ratios for azulene, obtained by dividing with the conductance or M-function of the 8,10 connectivity. The other rows show comparisons with experimental results from the literature and with our experimental results for pyrene. It is interesting to note that the mean-square deviations ($\chi^2$) of the MRR and GW predictions from the experimental azulene data are 0.37 and 0.44 respectively, which reveals that despite its simplicity, the mid-gap MRR is comparable with the accuracy of the GW calculation.

| Molecular heart | Anchor group | Connectivities | Literature notation | Mid-gap MRR | Experimental ratios | GW prediction ref [38] | DFT Prediction |
|---|---|---|---|---|---|---|---|
| Azulene | thiochroman | 4,9/8,10 | Ratio of molecules **2,6,AZ** and **1,3,AZ** of ref [38] | 0.72 | 1 | 0.32 | 0.93 |
| Azulene | thiochroman | 6,3/8,10 | Ratio of molecules **4,7,AZ** and **1,3,AZ** of ref [38] | 0.79 | 0.25 | 0.32 | 0.13 |
| Azulene | thiochroman | 3,5/8,10 | Ratio of molecules **5,7,AZ** and **1,3,AZ** of ref [38] | 0.003 | 0.06 | 0.1 | 0.05 |
| Naphthalene | thiol | 7,10/3,8 | Ratio of molecules **4** and **6** of ref [15] | 4 | 5.1 | Not available | 2 |
| Anthracene | thiol | See SI | Ratio of molecules **5** and **7** of ref [15] | 16 | 10.2 | Not available | 13 |
| Pyrene | carbon | 2,9/3,10 | **P1** and **P2** of this paper | 9 | 8 | Not available | 9 |
| Anthanthrene | pyridyl | See SI | Ratio of molecules **1** and **2** of ref [2] | 81 | 79 | Not available | 81 |

## PHASE-COHERENT INTERFEROMETERS AND LOGIC GATES

The MRR is derived under the assumption that transport through a molecule is phase coherent and since the agreement in Table 1 between theory and experiment suggests that this assumption is correct, it is natural utilise M-tables in the design of devices with more complex connectivities. In what follows, we examine theoretical concepts underpinning phase-coherent logic gates and transport through three-terminal devices, which illustrate the significance of the signs of the M-table entries.

In a three-terminal device with phase-coherent inputs of amplitudes $A_j$ and $A_k$, and an output site $j'$, the electrical conductance $\sigma_{jk;j'}$ is proportional to $|A_j M_{jj'} + A_k M_{kj'}|^2$ rather than $|A_j M_{jj'}|^2 + |A_k M_{kj'}|^2$. This leads to strategies for designing phase-coherent transistors and logic gates and optimising the sensitivity of molecular-scale Aharonov-Bohm interferometers. As an example, Figure 5a shows a pyrene-based XOR gate, whose truth table (figure 5b) is obtained from the fact that from the M-table of figure 3b, $M_{4,9} = - M_{2,9} = 3$. Clearly higher order logic gates could be obtained by combining elementary functions such as these. If the core of a molecule is gated by a third electrode, such that $E_F$ no longer coincides with the mid-gap, then the signs of M-functions at non-mid-gap energies are relevant and the electrical conductance is proportional to $\sigma_{jk;j'}(E_F) = |M_{jj'}(E_F) + M_{kj'}(E_F)|^2$.

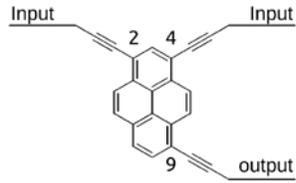

| Input | | Output |
|---|---|---|
| A1 | A2 | $|A1+A2|^2$ |
| 0 | 0 | 0 |
| 1 | 0 | 1 |
| 0 | 1 | 1 |
| 1 | 1 | 0 |

**Fig. 5** A pyrene-based phase-coherent XOR gate. (a) connections 2,9 and 4,9 with M numbers of opposite sign and equal magnitude (b) the resulting truth table. (c) The Fermi-energy dependence of $M_{29}$ (blue), $M_{49}$ (red), and $\sigma_{2,4;9}$ (green) (d) The Fermi-energy dependence of $M_{29}$ (blue), $M_{69}$ (red), and $\sigma_{2,6;9}$ (green). All quantities are plotted against the dimensionless Fermi energy $E_M^F = (E_F - E_{HL})/\delta_{HL}$.

For the three-fold connectivity of figure 5a, figure 5c,d shows a plot of the electrical conductances $\sigma_{2,4;9}(E_F)$ and $\sigma_{2,6;9}(E_F)$ versus $E_F$ and demonstrates that M-functions provide insight into the gate dependence of multiply-connected cores and can be used to select connectivities, which enhance or reduce the sensitivity to electrostatic gating. Since reproducible three-terminal devices are not currently available in the laboratory, we illustrate the use of M-functions in three-terminal devices through a theoretical study of molecular-scale Aharonov-Bohm (A-B) effect and a molecular-scale logic gate.

A schematic of an A-B device in which a top electrode is connected to a metallic loop, through which a magnetic flux is passed is shown in Fig. 6a. The loop connects to two sites $j$ and $k$ and the current exits through a bottom electrode connected to site $j'$. For incoming waves whose amplitudes differ only by a phase $\theta$, mid-gap electrical conductance is proportional to $\sigma_{jk;\,j'} = |M_{jj'}(E_{HL}) + e^{i\theta}M_{kj'}(E_{HL})|^2$. If $j$, $k$ and $j'$ are chosen to be 2, 6 and 13, then since $M_{2,13}(E_{HL}) = -3$ and $M_{6,13}(E_{HL}) = -1$, this yields $\sigma_{2,6;13} = |-3 + -1e^{i\theta}|^2 = 10 + 6\cos(\theta)$ and therefore the amplitude of oscillation (12) is 12/16 = 75% of the maximum value. The green curve of figure 6 shows the result of a complete tight-binding calculation of the conductance versus magnetic flux through the loop $\varphi$ in units of the flux quantum $\varphi_0$. This is related to the phase $\theta$ by $\theta = 2\pi\varphi/\varphi_0$.

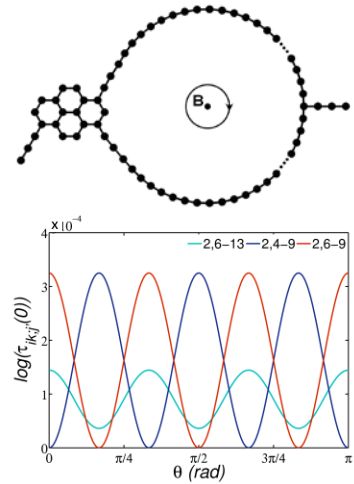

**Fig. 6** (a) Schematic of an Aharonov–Bohm loop in which the two arms of the loop are connected to different atoms of the pyrene core. A magnetic field B creates a flux $\varphi$ through the loop and a relative phase shift $\theta = 2\pi\varphi/\varphi_0$ for partial de Broglie waves traversing the different arms. (b) The electrical conductance as a function of the dimensionless flux $\theta$ for the different connection points to the pyrene core. The largest amplitude (ie flux sensitivity) occurs for the connectivities $\sigma_{2,4;9}$ and $\sigma_{2,6;9}$.

To illustrate how M-tables can be utilized in improving the sensitivity of such interferometers, we now seek to increase this amplitude to 100%. From the M-table, the solution is immediately obvious, because $M_{2,9}(E_{HL}) = -3$ and $M_{6,9}(E_{HL}) = -3$, and therefore if $j$, $k$ and $j'$ are chosen to be 2, 6 and 9, one obtains $\sigma_{2,6;9} = |-3 - 3e^{i\theta}|^2 = 18 + 18\cos(\theta)$, yielding a 100% amplitude. On the other hand since $M_{2,9}(E_{HL}) = -3$ and $M_{4,9}(E_{HL}) = +3$, if $j$, $k$ and $j'$ are chosen to be 2, 4 and 9, one obtains $\sigma_{2,4;9} = |-3 - 3e^{i\theta}|^2 = 18 - 18\cos(\theta)$ and therefore a π-shifted interferometer with a 100% amplitude is obtained. These features are demonstrated by performing a tight binding calculation (see methods) of the structure of figure 6a, the result of which is shown in figure 6b.

## CONCLUSION

When electrons enter the heart of a PAH at site $j$, then provided the coupling to the linkers is sufficiently weak, the amplitude of the resulting de Broglie wave at site $i$ is proportional to the M-function $M_{ij}(E)$. Although the associated electrical conductance $\sigma_{ij}$ depends on the nature of the coupling to the electrodes, the ratio of two such conductances with different choices of $ij$ does not. We have shown that mid-gap M-functions correctly predict conductance ratios of molecules with bipartite cores such as pyrene and non-bipartite cores such as azulene. Despite the simplicity of this parameter-free theory, quantitative agreement with experiment and with density functional and many-body GW calculations was obtained. One of the reasons for this agreement is that the MRR is independent of the energy gap of the molecules. Therefore even though a nearest-neighbour tight binding model may not be capable of describing the band gap for some large PAHs, where hydrogen-induced edge distortion is important (see for instance [47]), conductance ratios are correctly predicted.

Since energy-dependent M-functions can be obtained from gap-centre M-tables, this agreement between gap-centre values and experiment gives us confidence that M-functions correctly predict the energy dependence of interference patterns and superpositions of these patterns in multiply-connected molecules. As demonstrations of their utility, we have shown that M-functions can be used to design phase-coherent logic gates and to optimise the sensitivity of molecular Aharonov-Bohm and electrostatically-gated interferometers. The concept of energy-dependent M-functions is general and in contrast with theories of transport resonances, focusses attention on the opposite limit of transport in the vicinity of the mid-gap. These functions are properties of a molecular core and generalise quantum interference rules to arbitrary energies within the H-L gap.

**Supporting Information**

The SI describes the relationship between M-functions and Greens functions and contains a summary of the properties of M-functions. Derivations of analytic formulae for Greens functions and M-functions of the molecules of fig 2 are presented. M-functions of these and a more complex tetracene-based "cross molecule" ar plotted. A method of computing finite energy M-functions from mid-gap M-tables is presented. An expression for the core transmission coefficient is derived. X-ray crystallographic data is provided.

**Acknowledgements**

This work was supported by the Swiss National Science Foundation (No. 200021-147143) as well as by the European Commission (EC) FP7 ITN "MOLESCO" project no. 606728 and UK EPSRC, (grant nos. EP/K001507/1, EP/J014753/1, EP/H035818/1).

**Computational Methods**

The Hamiltonian of the structures described in this paper was obtained using DFT as described below or constructed from a simple tight-binding model with a single orbital per atom of site energy $\varepsilon_0 = 0$ and nearest-neighbor couplings $\gamma = -1$. To calculate the Aharonov–Bohm effect, for the different connection points to the pyrene (fig. 6), Aharonov–Bohm loop is a ring of 11 atoms with $\varepsilon_{AB\text{-}loop} = 0.05$ and $\gamma_{AB\text{-}loop} = -e^{i\theta}$ where $\theta$ varies in the interval of $[0, \pi]$.

*1. DFT calculation:* The optimized geometry and ground state Hamiltonian and overlap matrix elements of each structure was self-consistently obtained using the *SIESTA*[48] implementation of density functional theory (DFT). *SIESTA* employs norm-conserving pseudo-potentials to account for the core electrons and linear combinations of atomic orbitals to construct the valence states. The generalized gradient approximation (GGA) of the exchange and correlation functional is used with the Perdew-Burke-Ernzerhof parameterization (PBE)[49] a double-$\zeta$ polarized (DZP) basis set, a real-space grid defined with an equivalent energy cut-off of 250 Ry. The geometry optimization for each structure is performed to the forces smaller than 40 meV/Å.

*2. DFT-NEGF Transport calculation:* The mean-field Hamiltonian obtained from the converged DFT calculation or a simple tight-binding Hamiltonian was combined with our implementation of the non-equilibrium Green's function method, the *GOLLUM*,[50] to calculate the phase-coherent, elastic scattering properties of the each system consisting of left (source) and right (drain) leads and the scattering region. The transmission coefficient $T(E)$ for electrons of energy $E$ (passing from the source to the drain) is calculated via the relation: $T(E) = Tr(\Gamma_R(E)G^R(E)\Gamma_L(E)G^{R\dagger}(E))$. In this expression, $\Gamma_{L,R}(E) = i\left(\sum_{L,R}(E) - \sum_{L,R}^{\dagger}(E)\right)$ describe the level broadening due to the coupling between left (L) and right (R) electrodes and the central scattering region, $\sum_{L,R}(E)$ are the retarded self-energies associated with this coupling and $G^R = (ES - H - \sum_L - \sum_R)^{-1}$ is the retarded Green's function, where $H$ is the Hamiltonian and $S$ is overlap matrix. Using the obtained transmission coefficient ($T(E)$), the conductance could be calculated[51] by Landauer formula ($G = G_0 \int dE\, T(E)(-\partial f/\partial E)$) where $G_0 = 2e^2/h$ is the conductance quantum and $f(E) = $ $(1 + exp((E - E_F)/k_B T))^{-1}$ is the Fermi-Dirac distribution function, $T$ is the temperature and $k_B = 8.6 \times 10^{-5}$ eV/K is the Boltzmann's constant.

**3. Analytical methods:** M-functions are related to the Green's function of an isolated core by $M_{i,j}(E)=D(E)\,G_{i,j}(E)$, where $D(E)$ is proportional to $det(E-H)$, divided by a polynomial to remove degenerate eigenvalues. Since we are only interested in sites $i,j$ which can be connected by linkers to external electrodes, we solve Dyson's equation to obtain the "peripheral Green's function" $G_{i,j}(E)$ connecting only sites on the periphery of a core.

# *Supplementary Information*

# Searching the hearts of graphene-like molecules for simplicity, sensitivity and logic.

Sara Sangtarash,‡[*a] Cancan Huang,‡[b] Hatef Sadeghi,‡ [a] Gleb Sorohhov,[b] Jürg Hauser,[b] Thomas Wandlowski,[b] Wenjing Hong,[*b] Silvio Decurtins,[b] Shi-Xia Liu[*b] and Colin J. Lambert[*a]

[a] Quantum Technology Centre, Lancaster University, Lancaster LA1 4YB, UK. [b] Department of Chemistry and Biochemistry, University of Bern, Freiestrasse 3, CH-3012 Bern, Switzerland. [*]s.sangtarash@lancaster.ac.uk; c.lambert@lancaster.ac.uk; hong@dcb.unibe.ch; liu@dcb.unibe.ch. ‡ These authors contributed equally to this work.

## 1. Relationship between M-functions and Greens functions.

An M-function M(E) is related to the corresponding Greens function $G(E)$ by $M(E) = D(E)G(E)$, where $D(E)$ is chosen to cancel any divergences in $G(E)$. For a molecular heart described by a Hamiltonian H, the corresponding Greens function is defined by $(E - H)G = 1$. ie $G(E) = F(E)/det(E - H)$, where $F(E)$ is the transpose of the cofactor matrix of $(E - H)$. If $H|\varphi_n\rangle = E_n|\varphi_n\rangle$, then $(E) = \sum_n \frac{|\varphi_n\rangle\langle\varphi_n|}{E-E_n}$, $det(E - H) = \prod_n (E - E_n)$ and $F(E) = G(E)\det(E - H)$.

Unlike $G$, which contains poles at the eigenvalues $E_n$ of H, $F(E)$ is analytic everywhere. To obtain an M-function, which is also analytic everywhere, we might be tempted to choose $M(E) = aF(E)$, where $a$ is an arbitrary constant. This is equivalent to choosing $M(E) = D(E)G(E)$, where $D(E) = a\, det(E - H)$. Such a choice yields

$$M(E) = aG(E)\det(E - H) = \sum_n A_n \frac{|\varphi_n\rangle\langle\varphi_n|}{E-E_n} \quad (1)$$

where $A_n = a \prod_{m\,(m \neq n)}(E_m - E_n)$. With this choice, if $E_H$ and $|\varphi_H\rangle$ define the HOMO, then provided the HOMO is non-degenerate, when $E$ coincides with the HOMO energy $E_H$, one obtains $M(E_H) = A_H|\varphi_H\rangle\langle\varphi_H|$ where $A_H = a \prod_{n, E_n \neq E_H}(E_H - E_n)$.

Hence writing $\langle i|\varphi_H\rangle = \varphi_i^H$, we find $M_{ij}(E_H) = A_H \varphi_i^H \varphi_j^{H*}$. Unlike $G_{ij}(E_H)$, which diverges at $E = E_H$, $M_{ij}(E_H)$ is finite and the dependence on *i* is the same for all *j*. In other words, the *i* dependence of $M_{ij}(E_H)$ reproduces the *i* dependence of $\varphi_i^H$. The same argument holds for a non-degenerate LUMO $E_L$, $|\varphi_L\rangle$ and therefore as $E$ varies from $E_H$ to $E_L$, the *i*-dependence of $M_{ij}(E)$ changes smoothly from

$\varphi_i^H$ at $E = E_H$ to a $j$-dependent interference pattern at intermediate $E$, to a $j$-independent pattern $\varphi_i^H$ at $E = E_L$.

On the other hand, if the HOMO and LUMO are degenerate, then this desirable behaviour is lost, because $A_H = A_L = 0$ and the above definition would yield $M_{ij}(E_H) = M_{ij}(E_L) = 0$. To avoid this feature, the M-function associated with $G$ may be defined by $M(E) = D(E)G(E)$, where $D(E) = a \det(E - H)/P(E)$, In this expression, $a$ is an arbitrary constant and $P(E) = \prod_m(E - E_m)$, where the terms $m$ are chosen to cancel any degenerate roots in $\det(E - H)$. For example for a benzene ring with nearest-neighbour couplings -1 (see below), $\det(E - H) = (E^2 - 4)(E^2 - 1)^2$. Since the root $(E^2 - 1)$ appears twice, we eliminate this degeneracy by choosing $P(E) = (E^2 - 1)$. Similarly for a ring of 12 sites (see below) $\det(E - H) = (E^2 - 4)(E^2 - 3)^2(E^2 - 1)^2 E^2$ and therefore we choose $P(E) = (E^2 - 3)(E^2 - 1)E$. As an example, if only the HOMO $E_H$ is doubly degenerate with degenerate eigenstates $|\varphi_H\rangle$ and $|\varphi_h\rangle$, then

$$M(E_H) = D(E_H)[|\varphi_H\rangle\langle\varphi_H| + |\varphi_h\rangle\langle\varphi_h|] \qquad (2)$$

where $D(E) = a \det(E - H)/(E - E_H)$ and

$$M_{ij}(E_H) = D'(E_H)[\varphi_i^H \varphi_j^{H*} + \varphi_i^h \varphi_j^{h*}] \qquad (3)$$

which shows that in the presence of such a degeneracy, an electron of energy $E_H$ entering the core at orbital $j$ creates an interference pattern $M_{ij}(E_H)$ whose $i$-dependence is a superposition of the two wavefunctions $\varphi_i^H$ and $\varphi_i^L$, weighted by $\varphi_j^{H*}$ and $\varphi_j^{L*}$ respectively.

In the above expression for $D(E)$, the choice of the constant $a$ is arbitrary. For cores whose Hamiltonians are proportional to bipartite connectivity matrices, with a mid-gap energy $E_{HL} = 0$ we shall adopt the convention of choosing $a$ such that the mid-gap values $M_{ij}(E_{HL})$ are integers.

## 2. Summary of the properties of M-functions

The above discussion leads to M-functions with the following properties:

1. Viewed as a function of $i$ for fixed $j$, $M_{i,j}(E)$ is the amplitude of the interference pattern on atomic orbital $i$ of a core, due to electrons entering the core at $j$, with energy $E$.
2. If the HOMO $\varphi^H_i$ (LUMO $\varphi^L_i$) of the isolated core is non-degenerate, then $M_{i,j}(E_H)$ is proportional to $\varphi^H_i \varphi^{H*}_j$ ($M_{i,j}(E_L)$ is proportional to $\varphi^L_i \varphi^{L*}_j$)
3. If the HOMO possesses two degenerate orbitals, $\varphi^H_i$ and $\varphi^h_i$ then $M_{i,j}(E_H)$ is proportional to $\varphi^H_i \varphi^{H*}_j + \varphi^h_i \varphi^{h*}_j$ and similarly for a doubly-degenerate LUMO.
4. $M_{i,j}(E)$ is related to the Green's function $G_{i,j}(E)$ of a core by

$$M_{i,j}(E) = D(E)\, G_{i,j}(E) \qquad (4)$$

In this expression, $D(E) = a \det(E - H)/P(E)$, where $H$ is the Hamiltonian of the isolated molecule.

5. As a consequence, M-functions are analytic everywhere and in the range $E_H < E < E_L$, can be approximated by low-order polynomials in $E_M$.
6. For molecules which can be represented by bipartite lattices whose HOMO is completely filled and $E_{HL}=0$, M-functions $M_{i,j}(E)$ are either odd or even functions of $E$. For the former, $M_{i,j}(0)=0$, whereas for the latter $M_{i,j}(0)$ is an integer.
7. When linker groups, which are weakly coupled to orbitals $i$ and $j$, are in contact with metallic electrodes whose Fermi energy $E_F$ lies at the mid-gap $E = E_{HL}$, the resulting (low-temperature) electrical conductance $\sigma_{i,j}$ is proportional to $[M_{i,j}(E_{HL})]^2$. Therefore the ratio of two such conductances (associated with links $i$ and $j$ or $l$ and $m$) is given by the mid-gap M-ratio rule (MRR):

$$\sigma_{i,j}/\sigma_{l,m} = [M_{i,j}(E_{HL})/M_{l,m}(E_{HL})]^2 \qquad (5)$$

More generally for arbitrary $E_F$ in the vicinity of the mid-gap, $\sigma_{i,j}$ is proportional to $[M_{i,j}(E_F)]^2$ and the ratio of two conductances is given by:

$$\sigma_{i,j}/\sigma_{l,m} = [M_{i,j}(E_F)/M_{l,m}(E_F)]^2 \qquad (6)$$

8. If the molecule is weakly coupled to the electrodes via atomic orbitals $i$ and $j$, then the transmission coefficient $T_{ij}(E_M)$ is proportional to 'isolated-core transmission coefficient' $\tau_{i,j}(E)$, where

$$\tau_{i,j}(E) = (M_{i,j}(E)/D(E))^2 = (G_{i,j}(E))^2 \qquad (7)$$

Plots of $\tau_{i,j}(E)$ versus $E$ may be of interest for comparing with other theories of electron transmission, but are much more complicated than plots of $M_{i,j}(E)$, because of the presence of the denominator *D(E)*.

9. Depending on the precise molecule of interest, it may be convenient to choose $D(E) = a\,det(E - H)$. If the latter is chosen, then we refer to the resulting function is an associated M function. A degeneracy of the HOMO (or LUMO) will be signalled by $M_{ij}(E_H) = 0$ (or $M_{ij}(E_L) = 0$) for all $ij$. This definition is convenient numerically, because it yields *M(E)=aF(E)*, where *F(E)* is the easily-computed transpose of the cofactor matrix of *(E-H)*.

In principle, there is no advantage in using M-functions rather than Greens functions, because knowledge of *M(E)* and *D(E)* allows us to reproduce *G(E)* and vice versa. However in practice, there are several advantages. For example, in contrast with *G(0),* for bipartite PAH lattices, $M_{ij}(0)$ can be chosen to be a table of integers, which makes it an easy-to-use design tool and emphasises that conductance ratios are simply the squared ratio of two integers. Furthermore, M-functions are entire, which means that in contrast with Greens functions, their Taylor expansions in $E_M$ are convergent everywhere. From a practical viewpoint, this means that a low-order expansion in powers of $E_M$ reproduces M-functions more accurately than Greens functions. As an example, consider a benzene ring, whose M-function (see equ. 3 of the main text) is exactly reproduced by an expansion up to only $E_M^3$, whereas the Greens function of a benzene ring requires an infinite series to reproduce it exactly. In what follows, we shall generate low-order Taylor expansions in $E_M$ of the form

$$M_{ij}(E) = M_{ij}(E_{HL}) + M_{ij}^{(1)} E_M + M_{ij}^{(2)} E_M^2 + \cdots \qquad (8)$$

which means that for a given molecule, in addition to the magic integer table $M_{ij}(E_{HL})$, *n* additional M tables will be required. We show below that the higher order tables $M_{ij}^{(1)}$, $M_{ij}^{(2)}$ etc can be generated from a knowledge of $M_{ij}(E_{HL})$ and $D(E_{HL})$ alone.

As a third advantage, M-functions evaluated at *E=E_H* and *E=E_L* are well-behaved and coincide with molecular orbitals. In contrast, Greens functions diverge at these energies. Of course, one could extract molecular orbitals from Greens functions with an appropriate limiting procedure, but then the result of such a procedure is an M-function, which further emphasises the appeal of M-functions.

## 3. Derivation of analytic formulae for Greens functions and M functions.

There are two cases to consider.

**Case A: Some sites are on the periphery and some are in the interior.**

Examples of such lattices are pyrenes and phenylenylium in figure 2 of the main text. In this case we shall refer to the periphery as subspace A and the interior as subspace B. The two subspaces are connected by a Hamiltonian sub-matrix $H_{AB}$ and Dysons equation $(E - H)G = 1$ takes the form

$$\begin{pmatrix} E - H_{AA} & -H_{AB} \\ -H_{BA} & E - H_{BB} \end{pmatrix} \begin{pmatrix} G_{AA} & G_{AB} \\ G_{BA} & G_{BB} \end{pmatrix} = \begin{pmatrix} 1 & 0 \\ 0 & 1 \end{pmatrix} \qquad (9)$$

To write down the solution of this equation, it is convenient to introduce Greens functions $g_A$ and $g_B$ of decoupled subspaces A and B respectively, (ie when $H_{AB} = H_{BA}^\dagger = 0$) defined by $g_A^{-1} = E - H_{AA}$ and $g_B^{-1} = E - H_{BB}$. Then the solution is of the form

$$G_{AA} = g_A + g_A H_{AB} G_{BB} H_{BA} g_A \qquad (10)$$

where

$$G_{BB} = (g_B^{-1} - H_{BA} g_A H_{AB})^{-1} \qquad (11)$$

In the above expressions, if the periphery contains $N$ sites and the interior R sites, then $g_A, H_{AA}, G_{AA}$ ($g_B, H_{BB}, G_{BB}$) are $N \times N$ ($R \times R$) matrices, whereas $H_{AB}, G_{AB}$ are $N \times R$) matrices. The matrix elements of $g_A$ are simply the Greens function elements of a ring of N sites, with nearest neighbour couplings $-\gamma$ and are given by

$$g_{ij}(E) = d(E)^{-1} \cos k(|j - i| - N/2) \qquad (12)$$

In this expression,

$$d(E) = 2\gamma \sin k \sin kN/2 \qquad (13)$$

where $k(E) = cos^{-1}[-E/2\gamma]$. When computing the inverse on the right hand side of equ. (10), it will be necessary to obtain an expression for the determinant $\Delta$ defined by:

$$\Delta = det(g_B^{-1} - H_{BA}g_AH_{AB}) = det(E - H_{BB} - H_{BA}g_AH_{AB}) \tag{14}$$

This allows us to compute the determinant of $(E - H)$, because the determinant of any such block matrix is given by

$$det(E - H) = det\begin{pmatrix} E - H_{AA} & -H_{AB} \\ -H_{BA} & E - H_{BB} \end{pmatrix} = \Delta \, det(E - H_{AA}) \tag{15}$$

Although we shall be mainly interested in the peripheral Greens function $G_{AA}$, which is obtained after computing the interior Greens function $G_{BB}$, it is interesting to note that the Greens function linking interior to peripheral sites is given by

$$G_{BA} = g_B H_{BA} G_{AA} = G_{BB} H_{BA} g_A \tag{16}$$

Finally, we note that if $H_{AB}$ connects peripheral sites $p= p_1$, $p_2$ etc to interior sites, then from equation (10), the $ij^{th}$ matrix element of $G_{AA}$ (denoted $G_{ij}(E)$, where $i,j$ are peripheral sites) is given by

$$G_{ij}(E) = g_{ij}(E) + \sum_{pp'}[g_{ip}(E)V_{pp'}(E)g_{p'j}(E)] \tag{17}$$

$$V_{pp'}(E) = (H_{AB}G_{BB}H_{BA})_{pp'} \tag{18}$$

To evaluate equation (15), we note that $det(E - H_{AA}) = S_N$, where $S_N$ is the determinant of a ring of $N$ sites. When $N$ is even, to obtain $S_N$, we write $L=N/2$ and define $x_n = (\cos n\pi/L)^2$. Since the determinant of a matrix is the product of its eigenvalues, we find for $L$ odd,

$$S_N = (E^2 - 4)\prod_{n=1}^{(L-1)/2}(E^2 - 4x_n)^2 \tag{19}$$

and for $L$ even,

$$S_N = E^2(E^2 - 4)\prod_{n=1}^{(L-2)/2}(E^2 - 4x_n)^2 \tag{20}$$

Examples of these formulae, for the lattices in figure 2 of the main text are:

$N=6$, $L=3$, (a benzene ring):

$$S_6 = (E^2 - 4)(E^2 - 1)^2 = 2(1 - E^2)d(E)$$, where $d(E)$ is given by equation (13).

$N=10$, $L=5$, (Useful for azulene or naphthalene):

$$S_{10} = (E^2 - 4)(E^2 - 4\cos^2\tfrac{\pi}{5})^2(E^2 - 4\cos^2\tfrac{2\pi}{5})^2$$

$N=12$, $L=6$, (Useful for phenalenylium):

$$S_{12} = (E^2 - 4)(E^2 - 3)^2(E^2 - 1)^2 E^2$$

$N=14$, $L=14$, (Useful for pyrene):

$$S_{14} = (E^2 - 4)(E^2 - 4\cos^2\tfrac{\pi}{7})^2(E^2 - 4\cos^2\tfrac{2\pi}{7})^2(E^2 - 4\cos^2\tfrac{3\pi}{7})^2$$

**Example 1.**

As an example, for the phenalenylium cation, where $p_1 = 1$, $p_2 = 5$, $p_3 = 9$,

$$g_B^{-1} - H_{BA}g_A H_{AB} = E - \alpha^2 \sum_{pp'}[g_{pp'}(E)] = E - \frac{3\alpha^2}{d(E)}[\cos 6k + 2\cos k] \quad (21)$$

Hence $V_{pp'}(E)$ is independent of $p$ and $p'$ and given by

$$V_{pp'}(E) = \frac{\alpha^2}{\Delta}, \text{ where } \Delta = E - \frac{3\alpha^2}{d(E)}[\cos 6k + 2\cos k] \quad (22)$$

To aid conversion between $E$ and $E_M$, we note that $E_H = -1\gamma$, $E_L = 0\gamma$, the gap centre is at $E_{HL} = -0.5\gamma$. Furthermore, to obtain $M_{ij}(E)$ from $G_{ij}(E)$, we note that $D(E) = aS_{12}\Delta$, where $a = 3/5$.

**Example 2.**

As a second example, for pyrene,

$$G_{BB} = (g_B^{-1} - H_{BA}g_A H_{AB})^{-1} \quad (23)$$

where $(g_B^{-1} - H_{BA}g_A H_{AB}) = \begin{pmatrix} E & \gamma \\ \gamma & E \end{pmatrix} - \alpha^2 \begin{pmatrix} a_{p_1} + a_{p_2} & b_{p_1} + b_{p_2} \\ a_{p_3} + a_{p_4} & b_{p_3} + b_{p_4} \end{pmatrix} = \begin{pmatrix} V_2 & -V_1 \\ -V_1 & V_2 \end{pmatrix}.$

In this equation, $a_i = g_{ip_1} + g_{ip_2}$ and $b_i = g_{ip_3} + g_{ip_4}$ and using the numbering convention of figure 2 of the main text, $p_1 = 1, p_2 = 5, p_3 = 8, p_4 = 12$. After taking advantage of symmetry it has been noted that $V_1$ and $V_2$ simplify to $V_1 = (2\alpha^2 b_{p_1} - \gamma)$ and $V_2 = (E - 2\alpha^2 a_{p_1})$. Hence

$$G_{BB} = \frac{1}{\Delta}\begin{pmatrix} V_2 & V_1 \\ V_1 & V_2 \end{pmatrix} \tag{24}$$

where $\Delta = V_2^2 - V_1^2$. Hence $V_{p_1p_1}(E) = V_{p_2p_2}(E) = V_{p_3p_3}(E) = V_{p_4p_4}(E) = V_{p_1p_2}(E) = V_{p_2p_1}(E) = V_{p_3p_4}(E) = V_{p_4p_3}(E) = \frac{\alpha^2 V_2}{\Delta}$ and $V_{p_1p_3}(E) = V_{p_2p_3}(E) = V_{p_3p_2}(E) = V_{p_3p_1}(E) = V_{p_1p_4}(E) = V_{p_4p_1}(E) = V_{p_2p_4}(E) = V_{p_4p_2}(E) = \frac{\alpha^2 V_1}{\Delta}$. With this notation, Dyson's equation yields

$$G_{ij}(E) = g_{ij}(E) + \frac{\alpha^2 V_1}{\Delta}[a_i b_j + b_i a_j] + \frac{\alpha^2 V_2}{\Delta}[a_i a_j + b_i b_j] \tag{25}$$

Finally, $\det(E - H) = S_{14} \Delta$. For pyrene, $E_H = -0.45\gamma$, $E_L = 0.45\gamma$, the gap centre is at $E_{HL} = 0$ and $D(E) = a\, d(E)\Delta/(E^2 - \gamma^2)$ where $a = 1/6$. In the above expressions, $\alpha$ and $\gamma$ are shown explicitly. In what follows, to obtain a parameter-free description based on connectivity alone, we choose $\alpha = \gamma = 1$.

**Case B: All sites are on the periphery**

Examples of such lattices are azulene and naphthalene. In this case, since all sites belong to the periphery space A, we drop the above A, B notation and write Dysons equation as $(E - H)G = (E - H_0 - H_1)G = 1$, where $H = H_0 + H_1$. In this equation, $H_0$ is the Hamiltionian of a ring of sites and $H_1$ contains the extra couplings -$\alpha$, examples of which are shown in figure 2 of the main text, for azulene and naphthalene. $H_1$ and $H$ are $NxN$ matrices, but if only $R$ peripheral sites $p = p_1, p_2$ etc are connected by $H_1$ then all elements of $H_1$ are zero, except a $RxR$ submatrix connecting sites $p$, which we denote $\bar{H}_1$. In the case of both azulene and naphthalene,

$$\bar{H}_1 = \begin{pmatrix} 0 & -\alpha \\ -\alpha & 0 \end{pmatrix} \tag{26}$$

To solve Dysons equation, we introduce the Greens function $g$ of the ring of $N$ sites, given by equation (12), which satisfies $(E - H_0)g = 1$ and the $RxR$ submatrix of $g$ containing only matrix elements con-necting sites $p$, which we denote $\bar{g}$. Then the solution to Dysons equation is $G = g + gVg$, which in component form has exactly the same structure as equation (19), except that in this case $V$ is a $RxR$ submatrix given by

$$V = (1 - \bar{H}_1 \bar{g})^{-1} \bar{H}_1 \tag{27}$$

When computing the above inverse, it will be necessary to obtain the determinant $\Delta$ defined by:

$$\Delta = det(1 - \bar{H}_1 \bar{g}) \tag{28}$$

This allows us to compute the determinant of $(E - H)$, because

$$det(E - H) = det(E - H_0 - H_1) = det(E - H_0) \det(1 - \bar{H}_1 \bar{g}) = S_N \Delta. \tag{29}$$

**Example 3.**

As an example, for naphthalene and azulene,

$$(1 - \bar{H}_1 \bar{g}) = \begin{pmatrix} 1 & 0 \\ 0 & 1 \end{pmatrix} - \begin{pmatrix} 0 & -\alpha \\ -\alpha & 0 \end{pmatrix} \begin{pmatrix} g_{p_1 p_1} & g_{p_1 p_2} \\ g_{p_2 p_1} & g_{p_2 p_2} \end{pmatrix} = \begin{pmatrix} -V_2 & V_1 \\ V_1 & -V_2 \end{pmatrix} \tag{30}$$

where

$$V_1 = \alpha g_{p_2 p_2}, \quad V_2 = -(1 + \alpha g_{p_2 p_1}) \tag{31}$$

Therefore $\Delta = V_2^2 - V_1^2$ and

$$V = \frac{\alpha}{\Delta} \begin{pmatrix} V_1 & V_2 \\ V_2 & V_1 \end{pmatrix} \tag{32}$$

Hence

$$G_{ij}(E) = g_{ij}(E) + \frac{\alpha V_1}{\Delta}\left[g_{ip_1}g_{p_1j} + g_{ip_2}g_{p_2j}\right] + \frac{\alpha V_2}{\Delta}\left[g_{ip_1}g_{p_2j} + g_{ip_2}g_{p_1j}\right] \tag{33}$$

Adopting the numbering convention in figure 2, for naphthalene, $p_1 = 1$ and $p_2 = 6$, whereas for azulene, $p_1 = 1$ and $p_2 = 7$. For naphthalene, $E_H = -0.62\,\gamma$, $E_L = 0.62\,\gamma$ and as expected, the gap centre is at $E_{HL} = 0$. For azulene, $E_H = -0.48\gamma$, $E_L = 0.4\gamma$ and the gap centre is at $E_{HL} = -0.04\gamma$.

For both molecules,

$$\Delta = (1 + g_{p_2p_1})(1 + g_{p_1p_2}) - g_{p_2p_2}g_{p_1p_1}. \tag{34}$$

The above analytic results for M-functions are mere examples and many more are accessible. Even for molecules for which analytic results are unwieldy, M-functions provide a convenient means of capturing the key features of interference patterns. As an example, consider the X-shaped tetracene-based molecule of figure S1b for which a selection of M-functions is shown in figure S1d. As expected, these are either even or odd functions of $E$ and the mid-gap values are integers. Despite the size and complexity of this molecule, it is clear that M-functions are simple and can be approximated by low-order polynomials in $E_M$.

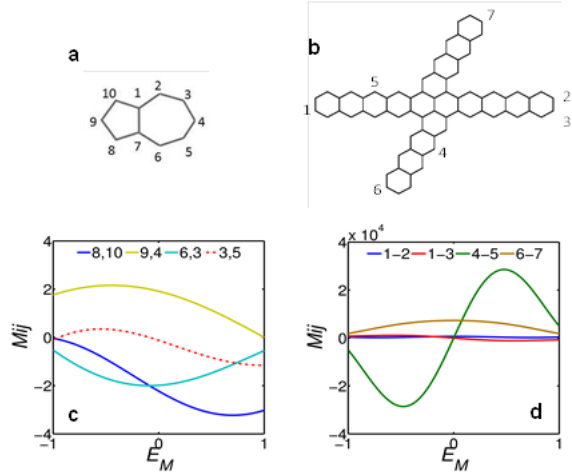

**Fig. S1** (a) Azulene molecular structure, (b) An X-shaped tetracene-based bipartite molecule, (c) Four M-functions of azulene, and (d) Selected M-functions for the X-shaped molecule b. Note that for the X-shaped molecule (plot d) the brown and blue curves are even functions of $E_M$ and the pink and green curves are odd functions of $E_M$. For azulene, the mid-gap M-functions evaluated at $E_M = 0$ are $M_{8,10}(0) = -0.55$, $M_{4,9}(0) = 0.47$, $M_{3,6}(0) = -0.49$ and $M_{3,5}(0) = -0.03$, which yields the ratios $(M_{4,9}(0)/M_{8,10}(0)) = 0.72$, $M_{3,6}(0)/M_{8,10}(0) = 0.79$ and $M_{3,5}(0)/M_{8,10}(0) = 0.003$.

## 4. The $\bar{C}$ and $\bar{M}(0)$ tables of the Naphthalene, Anthracene and Anthanthrene

For these bipartite lattices, zero-energy M-table $M(0)$ is block off-diagonal of the form

$$M(0) = \begin{pmatrix} 0 & \bar{M}^t \\ \bar{M} & 0 \end{pmatrix}, \tag{35}$$

and

$$C = \begin{pmatrix} 0 & \bar{C}^t \\ \bar{C} & 0 \end{pmatrix} \tag{36},$$

where $\bar{M}$ is a table of integers.

| $\bar{C}$ | 1 | 2 | 3 | 4 | 5 |
|---|---|---|---|---|---|
| 1' | -1 | -1 | 0 | 0 | 0 |
| 2' | 0 | -1 | -1 | 0 | 0 |
| 3' | 0 | 0 | -1 | -1 | 0 |
| 4' | 0 | -1 | 0 | -1 | -1 |
| 5' | -1 | 0 | 0 | 0 | -1 |

| $\bar{M}$ | 1' | 2' | 3' | 4' | 5' |
|---|---|---|---|---|---|
| 1 | -2 | 1 | -1 | 1 | -1 |
| 2 | -1 | -1 | 1 | -1 | 1 |
| 3 | 1 | -2 | -1 | 1 | -1 |
| 4 | -1 | 2 | -2 | -1 | 1 |
| 5 | 2 | -1 | 1 | -1 | -2 |

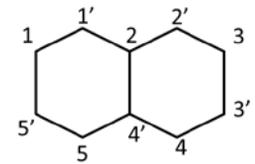

naphthalene

| $\bar{C}$ | 1 | 2 | 3 | 4 | 5 | 6 | 7 |
|---|---|---|---|---|---|---|---|
| 1' | -1 | -1 | 0 | 0 | 0 | 0 | 0 |
| 2' | 0 | -1 | -1 | 0 | 0 | 0 | 0 |
| 3' | 0 | 0 | -1 | -1 | 0 | 0 | 0 |
| 4' | 0 | 0 | 0 | -1 | -1 | 0 | 0 |
| 5' | 0 | 0 | -1 | 0 | -1 | -1 | 0 |
| 6' | 0 | -1 | 0 | 0 | 0 | -1 | -1 |
| 7' | -1 | 0 | 0 | 0 | 0 | 0 | -1 |

| $\bar{M}$ | 1' | 2' | 3' | 4' | 5' | 6' | 7' |
|---|---|---|---|---|---|---|---|
| 1 | -3 | 1 | -1 | 1 | -1 | 1 | -1 |
| 2 | -1 | -1 | 1 | -1 | 1 | -1 | 1 |
| 3 | 1 | -1 | -1 | 1 | -1 | 1 | -1 |
| 4 | -1 | 1 | -3 | -1 | 1 | -1 | 1 |
| 5 | 1 | -1 | 3 | -3 | -1 | 1 | -1 |
| 6 | -1 | 1 | -1 | 1 | -1 | -1 | 1 |
| 7 | 3 | -1 | 1 | -1 | 1 | -1 | -3 |

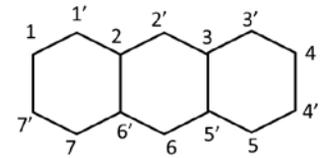

anthracene

| $\bar{M}$ | 1' | 2' | 3' | 4' | 5' | 6' | 7' | 8' | 9' | 10' | 11' |
|---|---|---|---|---|---|---|---|---|---|---|---|
| 1 | -9 | 7 | -4 | 4 | -1 | 1 | -1 | 1 | -1 | 2 | -3 |
| 2 | -1 | -7 | 4 | -4 | 1 | -1 | 1 | -1 | 1 | -2 | 3 |
| 3 | 1 | -3 | -4 | 4 | -1 | 1 | -1 | 1 | -1 | 2 | -3 |

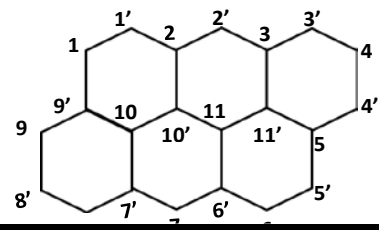

| | | | | | | | | | | |
|---|---|---|---|---|---|---|---|---|---|---|
| 4 | -1 | 3 | -6 | -4 | 1 | -1 | 1 | -1 | 1 | -2 | 3 |
| 5 | 1 | -3 | 6 | -6 | -1 | 1 | -1 | 1 | -1 | 2 | -3 |
| 6 | -1 | 3 | -6 | 6 | -9 | -1 | 1 | -1 | 1 | -2 | 3 |
| 7 | 3 | -9 | 8 | -8 | 7 | -7 | -3 | 3 | -3 | 6 | 1 |
| 8 | -6 | 8 | -6 | 6 | -4 | 4 | -4 | -6 | 6 | -2 | -2 |
| 9 | 6 | -8 | 6 | -6 | 4 | -4 | 4 | -4 | -6 | 2 | 2 |
| 10 | 3 | 1 | -2 | 2 | -3 | 3 | -3 | 3 | -3 | -4 | 1 |
| 11 | -2 | 6 | -2 | 2 | 2 | -2 | 2 | -2 | 2 | -4 | -4 |

anthanthrene

# 5. Plots of selected M-functions

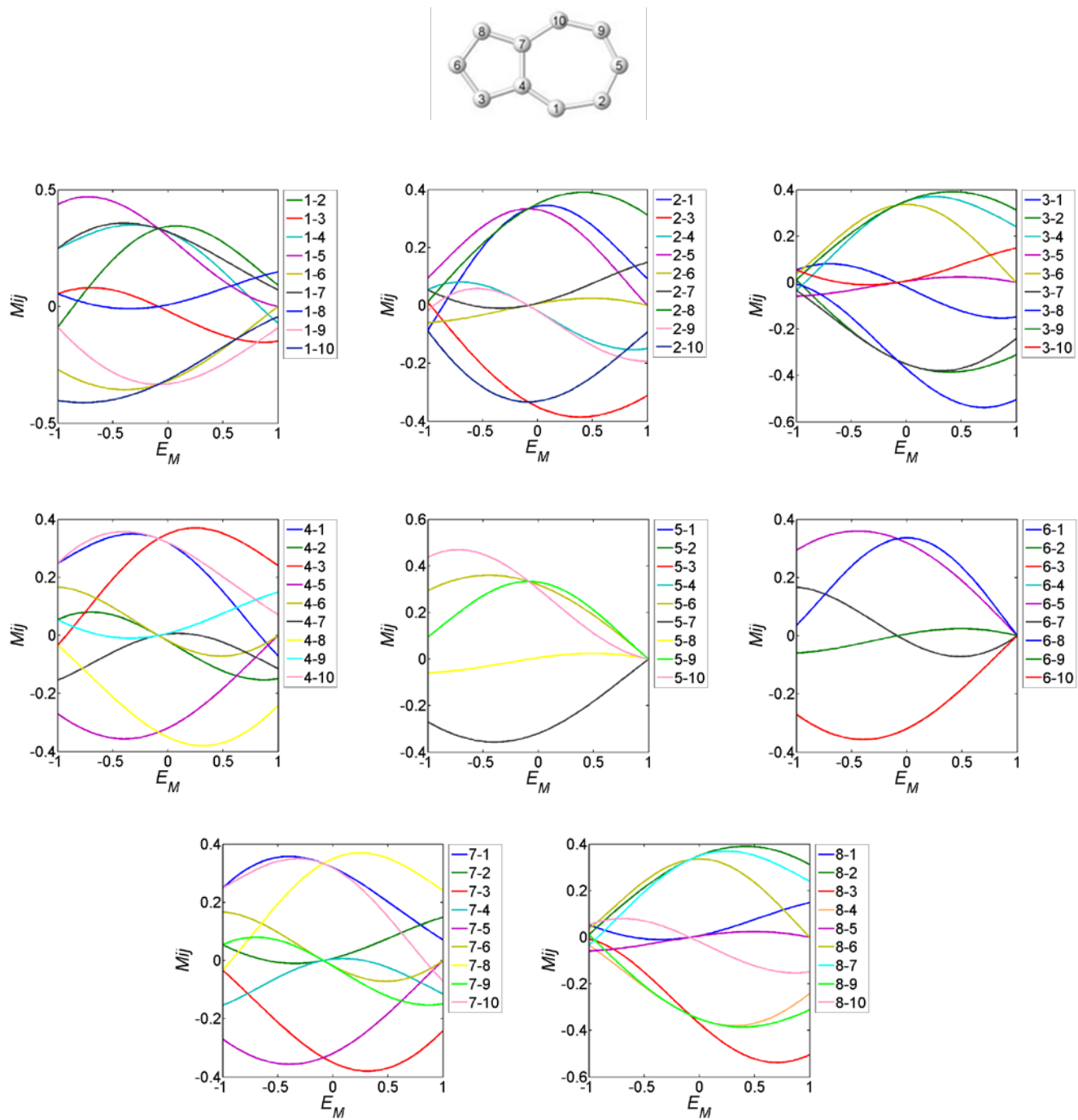

**Fig. S2** M functions for Azulene

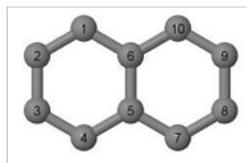

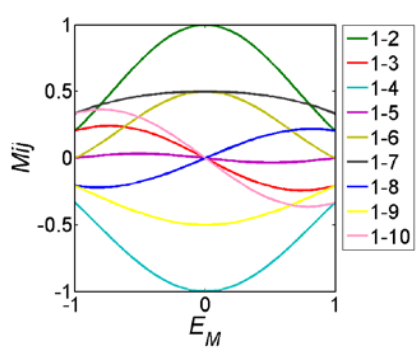
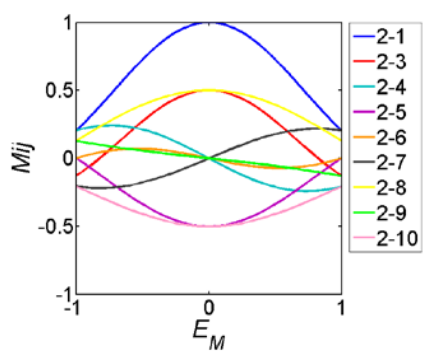
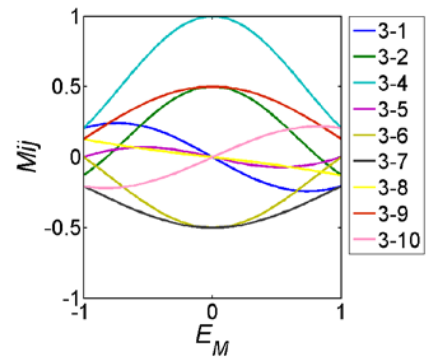

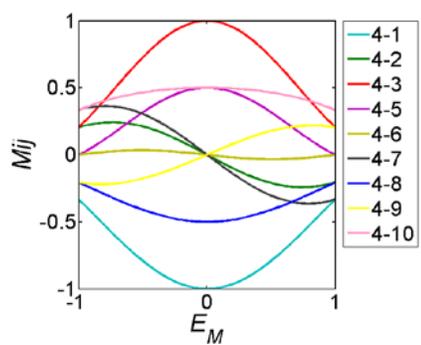
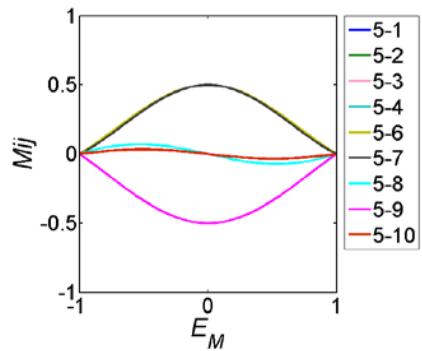

**Fig. S3** M functions for Naphthalene

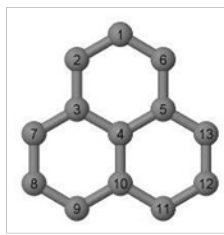
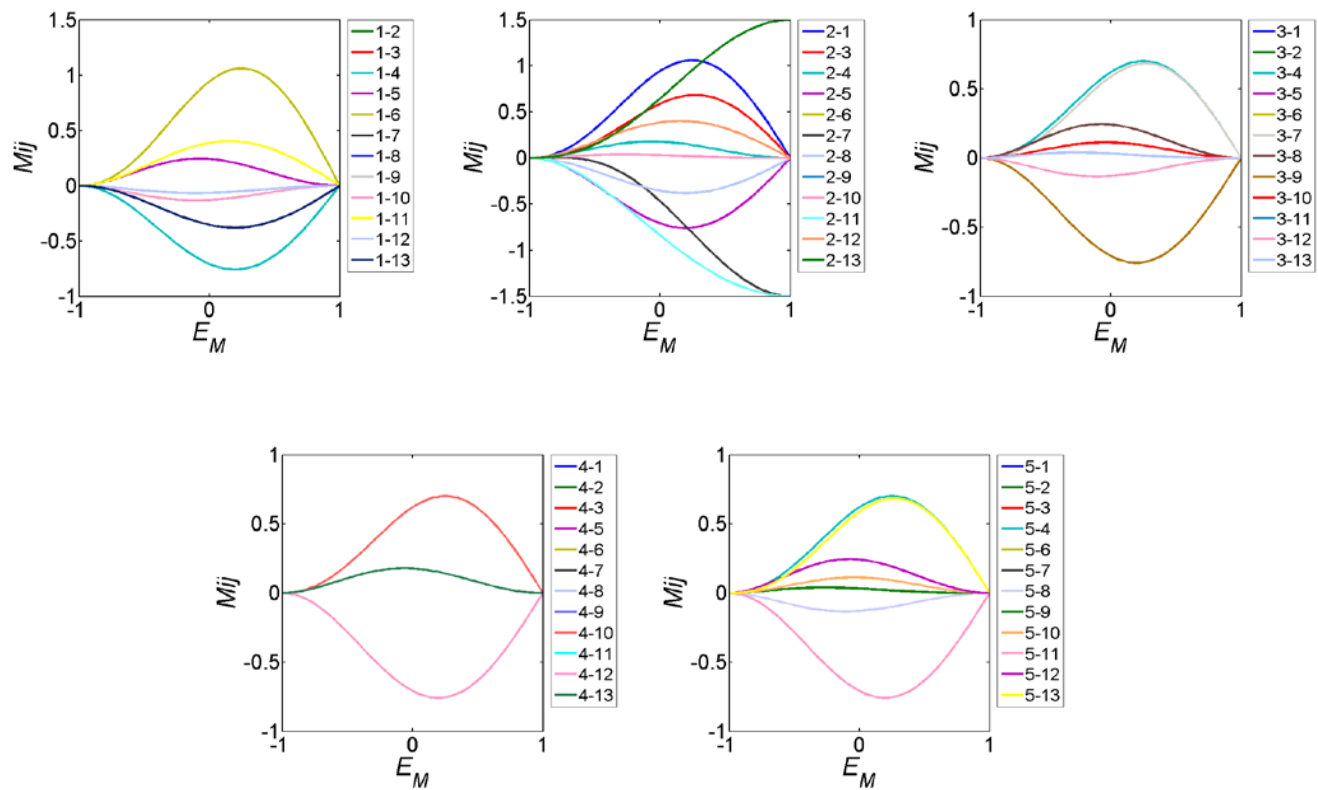

**Fig. S4** M functions for phenalenylium cation

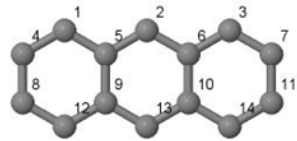

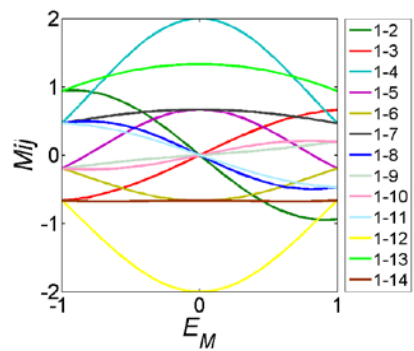
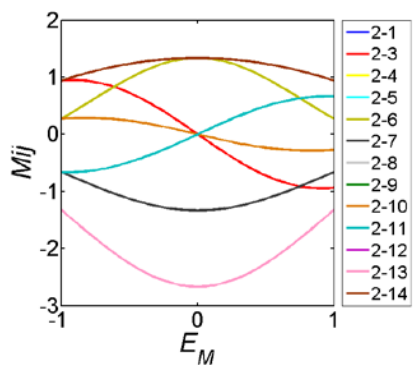
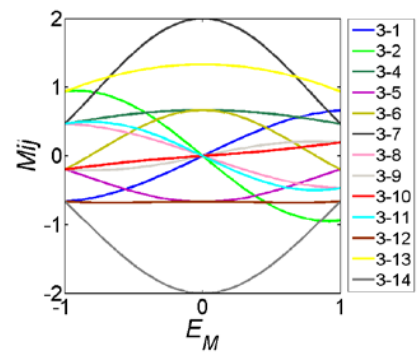

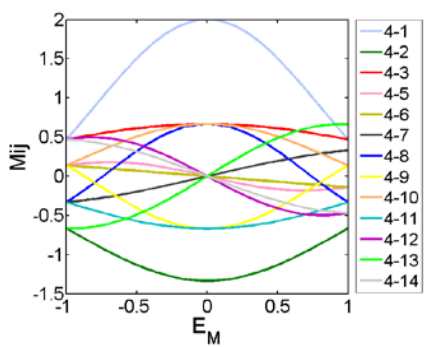
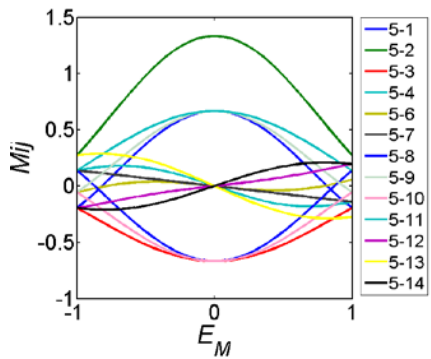
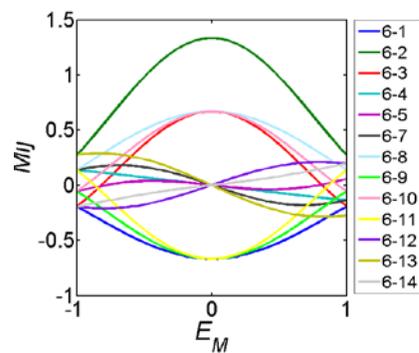

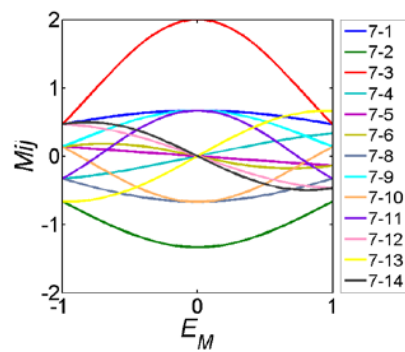

**Fig. S5** M functions for Anthracene

## 6. Isolated-core transmission coefficients for azulene: comparison with GW theory

Table 1 of the main text illustrates that the MRR correctly predicts experimental trends. From a theoretical viewpoint, it is also of interest to check that this analytical theory predicts the energy dependence of transmission coefficients, which (via M-function property 8) are proportional to the core transmission coefficient $\tau_{ij}(E) = |M_{ij}(E)/D(E)|^2 = |G_{ij}(E)|^2$. Plots of these functions are shown in the figure below. These are in qualitative agreement with the results of large-scale many-body GW calculations presented in ref [1], which is again remarkable given the simple nature of our "M-theory". Clearly graphs of $\tau_{ij}(E)$ are much more complicated than graphs of $M_{ij}(E)$, because of the presence of the denominator $D(E)$, which from the viewpoint of conductance ratios, is irrelevant.

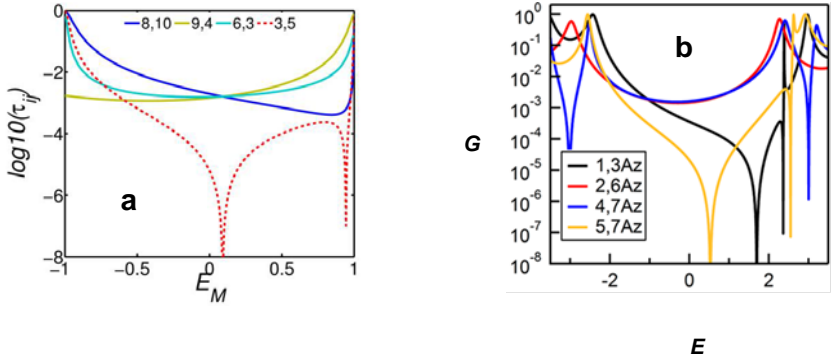

**Fig. S6** (a) Four transmission coefficients $\tau_{ij}(E_M)$ of the azulene core (d) The four GW transmission coefficients of the azulene reproduced from ref[1].

## 7. Taylor expansions of M-functions, Greens functions and core transmission functions.

In terms of the mid-gap energy $E_{HL}$,

$G(E) = (E - H)^{-1} = (E_{HL} - H)^{-1}[1 + (E - E_{HL})(E_{HL} - H)^{-1}]^{-1} = (E_{HL} - H)^{-1}\sum_{n=0}^{\infty}(E_{HL} - H)^{-n}[-(E - E_{HL})]^n = G(E_{HL})\sum_{n=0}^{\infty}G(E_{HL})^{-n}[-(E - E_{HL})]^n$  (37)

Hence to order $(E - E_{HL})^2$,

$$G(E) = G(E_{HL}) + (E_{HL} - E)G^2(E_{HL}) + (E_{HL} - E)^2 G^3(E_{HL}) \qquad (38)$$

Equation (38) allows us to generate a low-order power series for M-functions, as follows: Since $M(E) = D(E)G(E)$, in what follows, it will be convenient to define $\varepsilon = (E - E_{HL})/D(E_{HL})$ and therefore to order $\varepsilon^2$,

$$G(E) = G(E_{HL})[1 - \varepsilon M(E_{HL}) + \varepsilon^2 M^2(E_{HL})] + \ldots \qquad (39)$$

and

$$M(E) = D(E)G(E_{HL})[1 - \varepsilon M(E_{HL}) + \varepsilon^2 M^2(E_{HL})] + \ldots \qquad (40)$$

In this expression, $D(E) = a\,det(E - H)/P(E)$, where $P(E)$ cancels degeneracies. To simplify this expression, we write

$$D(E) = D(E_{HL})(1 + \alpha\varepsilon + \beta\varepsilon^2 + \cdots) \qquad (41)$$

Hence

$$M(E) = M(E_{HL})\{1 + \varepsilon[\alpha - M(E_{HL})] + \varepsilon^2[\beta + M^2(E_{HL}) - \alpha M(E_{HL})]\} + \ldots \qquad (42)$$

To obtain $\alpha$ and $\beta$ we note that in general to order $\varepsilon^2$,

$$det(E - H) = det(E_{HL} - H)(1 + \alpha'\varepsilon + \beta'\varepsilon^2) \qquad (43)$$

Where $\alpha' = TrM(E_{HL})$ and $\beta' = \frac{1}{2}[(\alpha')^2 - Tr\,M^2(E_{HL})]$. Hence if

$$P(E) = (1 + p_1\varepsilon + p_2\varepsilon^2 + \cdots) \qquad (44)$$

so that $1/P(E) = (1 - p_1\varepsilon + \varepsilon^2(p_1^2 - p_2) + \cdots)$ then $D(E) = a\,det(E - H)/P(E)$ is given by equation (40), with

$$\alpha = \alpha' - p_1 \text{ and } \beta = \beta' + p_1^2 - p_2 - \alpha' p_1 \qquad (45)$$

Equation (42) shows how to generate energy-dependent M tables from the mid-gap table $M_{ij}(\mathbf{0})$. As an example for a benzene ring, $\mathbf{E_{HL}} = \mathbf{0}$, $a=1/2$, $\mathbf{P(E)} = (\mathbf{1} - \mathbf{E^2})$, $Tr\,M(0)=0$, $\mathbf{Tr\,M^2(0)} = \mathbf{18}$ and $\mathbf{det(E_{HL} - H)} = -\mathbf{4}$, $\mathbf{D(E_{HL})} = -\mathbf{2}$, $\boldsymbol{\alpha} = \boldsymbol{\alpha'} = \mathbf{0}$, $\mathbf{p_1} = \mathbf{0}$, $\mathbf{p_2} = -\mathbf{D^2(E_{HL})} = -\mathbf{4}$. This yields

$$\mathbf{M(E)} = \mathbf{M(0)}\{\mathbf{1} + \frac{E}{2}\mathbf{M(0)} + \frac{E^2}{4}[\mathbf{M^2(0)} - \mathbf{5}]\} + \cdots \qquad (46)$$

Utilising the matrix elements $M_{ij}(E)$ in the MRR will give a more accurate approximation to conductance ratios than utilising $G_{ij}(E)$. Nevertheless for the purpose of computing energy integrals of transmission functions, it will be necessary to compute integrals of squared matrix elements of the form $\tau_{ij}(E) = [G_{ij}(E)]^2$.

To obtain expressions for such integrals, we note that to order $(E - E_{HL})^2$,

$$G_{ij}(E) = G_{ij}(E_{HL}) - (E - E_{HL})[G^2(E_{HL})]_{ij} + (E - E_{HL})^2[G^3(E_{HL})]_{ij} \qquad (47)$$

In the above expression, $[G^2(E_{HL})]_{ij}$ is obtained by first computing the squared matrix $G^2(E_{HL})$ and then taking the $ij$th element of the result.

To simplify the notation, it is useful to write $G(E_{HL}, n) = G(E_{HL})^n$, so that the above expression becomes

$$G_{ij}(E) = G_{ij}(E_{HL}, 1) - (E - E_{HL})G_{ij}(E_{HL}, 2) + (E - E_{HL})^2 G_{ij}(E_{HL}, 3) \qquad (48)$$

This yields for the core transmission coefficient,

$$\tau_{ij}(E) = [G_{ij}(E)]^2 = \tau_{ij}^{(0)} + (E_M)\tau_{ij}^{(1)} + (E_M)^2 \tau_{ij}^{(2)} \qquad (49)$$

where

$$\tau_{ij}^{(0)} = [G_{ij}(E_{HL})]^2, \qquad (50)$$

$$\tau_{ij}^{(1)} = -2\delta_{HL} G_{ij}(E_{HL}) G_{ij}(E_{HL}, 2) \qquad (51)$$

and

$$\tau_{ij}^{(2)} = [\delta_{HL}]^2[(G_{ij}(E_{HL}, 2))^2 + 2G_{ij}(E_{HL}) G_{ij}(E_{HL}, 3)^3]. \qquad (52)$$

## 8. The HOMO and LUMO energies and mid gap energies of the molecules

|  | $E_{LUMO}$ | $E_{HOMO}$ | $E_{HL}$ |
|---|---|---|---|
| **Pyrene** | 0.4450 | -0.4450 | 0 |
| **Naphthalene** | 0.6180 | -0.6180 | 0 |
| **Phenalenylium cation** | 1.0000 | 0 | 0.5 |
| **Azulene** | 0.4004 | -0.4773 | -0.038 |
| **Benzene** | 1.0000 | -1.0000 | 0 |
| **Anthracene** | 0.4142 | -0.4142 | 0 |
| **Tetracene tetramer** | 0.1953 | -0.1953 | 0 |

## 9. X-ray

X-ray crystallographic data for **P1** in CIF format; CCDC 1051154.

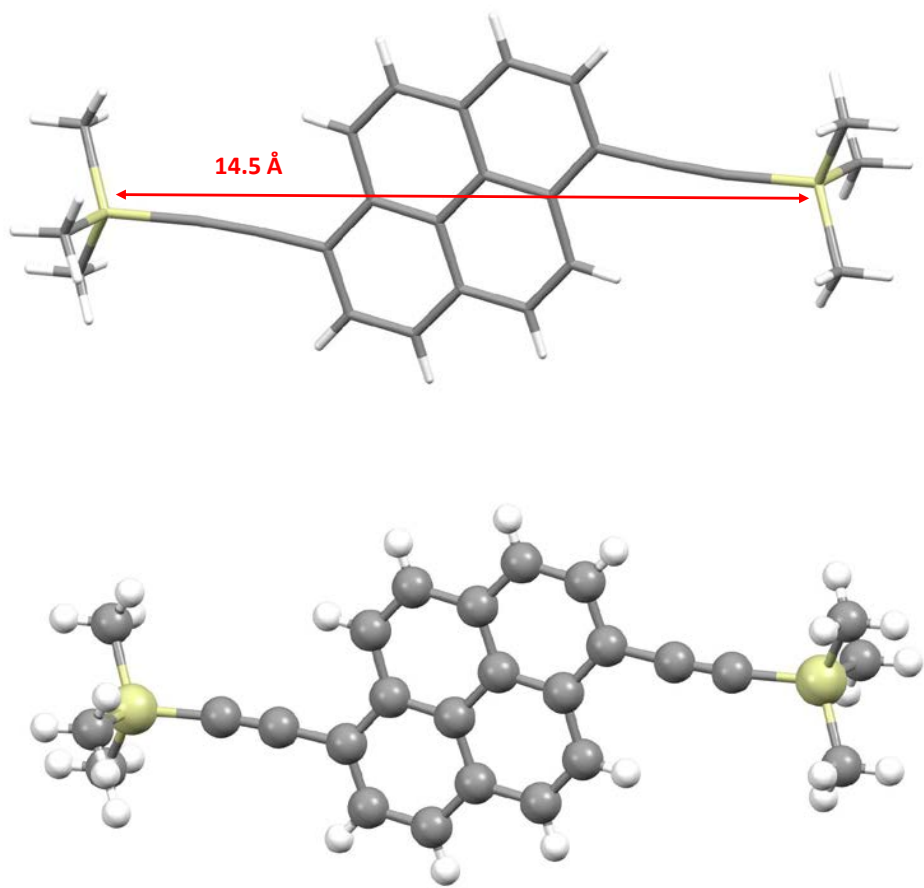

**Fig. S7** Molecular structure of **P1** (see also Fig. 1, main text) with the Si···Si distance indicated, which after the desilylation reaction and trapping of the molecule (Au − C bond formation) corresponds to the Au − **P1** − Au separation (top: capped stick presentation; bottom: ball and stick representation).

**SI References**